# Alkalized Borazine: A Simple Recipe to Design Superalkali Species

Ambrish Kumar Srivastava and Neeraj Misra[*]

*Department of Physics, University of Lucknow, Lucknow- 226007, India*

[*]Corresponding author's E-mail: neerajmisra11@gmail.com

**Abstract**

We propose a simple yet effective route to the design of superalkalis; by successive alkali metal substitution in borazine ($B_3N_3H_6$). Using Li atoms, our density functional calculations demonstrate that the vertical ionization energy (VIE) of $B_3N_3H_{6-x}Li_x$ decreases with the increase in *x* for *x* = 1 to 6. For *x* ≥ 4, the VIE of $B_3N_3H_{6-x}Li_x$ becomes lower than that of Li atom, thereby indicating their superalkali nature. More interestingly, all these species are planar such that $NICS_{zz}$ value at the ring's center is reduced. These novel superalkalis are expected to stimulate further interests in this field.

## 1. Introduction

During 1980s, the hypervalent species possessing lower ionization energy than alkali metals had been reported as superalkali species by Gutsev and Boldyrev [1]. They suggested a general formula of $XM_{n+1}$ for superalkalis, where M is alkali atom and X is electronegative atom with valence *n*. Typical examples of superalkalis include $FLi_2$, $OLi_3$, $NLi_4$ etc. Note that all these species contain nine valence electrons violating, at least formally, the octet rule. $OLi_3$ superalkali has been observed experimentally for the first time by Wu et al. [2] and $FLi_2$ superalkali has also been studied experimentally [3-6]. Superalkalis possess potential reducing capability and can be used in the synthesis of a variety of charge transfer salts and species with unusual properties. For instance, superalkalis can be employed to design supersalts with tailored properties [7-10], alkalides with anionic alkali metals [10-12] and superbases with strong basicity [13, 14]. Recently, $OLi_3O^-$ anion has been predicted to be the strongest base to date, whose proton affinity exceeds to that of $LiO^-$ [15]. Due to intriguing properties of superalkalis and their compounds, such species have been continuously explored [16-20]. Hou et al. [17] have reported the existence the non metallic superalkali cations such as $F_2H_3^+$, $O_2H_5^+$, $N_2H_7^+$ and $C_2H_9^+$. The binuclear superalkali cations belonging to the formula $M_2Li_{2k+1}$, where *k* is valence of M = F, O, N and C have been investigated by Tong et al. [18]. In a subsequent report [19], the authors have designed polynuclear superalkalis $YLi_3^+$ using various functional groups (Y= $CO_3$, $SO_3$, $SO_4$ etc.) as central core. The polynuclear superalkalis based on alkali-monocyclic (pseudo)-oxocarbon have been also reported [20]. The non-planar aromatic superalkalis have also been designed using inorganic such as $Be_3^{2-}$, $B_3^-$, $Al_4^{2-}$ etc. as well as organic $C_4H_4^{2-}$, $C_5H_5^-$ aromatic anions [21].

Borazine ($B_3N_3H_6$) is planar and isoelectronic to benzene, commonly referred to as inorganic benzene [22, 23]. Although it shows a different reactivity pattern due to presence of polar B−N bonds, many substitutions reactions of gaseous borazine have been reported [24, 25]. In this letter, we report an effective yet simple route to the design of superalkali species, by successive substitution of alkali metals in borazine. We show that successive alkalization tends to lower the ionization energy of the ring below alkali metals, consequently leads to a new class of planar superalkalis.

## 2. Computational details

All structures considered in this study were fully optimized at B3LYP method [26, 27] using correlation consistent basis set aug-cc-pVDZ [28] in Gaussian 09 program [29]. The geometry optimization was performed without any symmetry constraints and followed by frequency calculations to ensure that the optimized structures correspond to true minima in the potential energy surface. The vertical ionization energy (VIE) of systems has been calculated by difference of total energy of optimized structure of neutral structures and corresponding cations:

$$\text{VIE} = E_{\text{cation}} - E_{\text{neutral}} \qquad \text{(at optimized geometry of neutral structure)}$$

In order to analyze the stability of ring structure by substitution of alkali metal atoms (M), we have calculated substitution energy ($E_s$) and its increment ($\Delta E_s$) using following formula:

$$E_s = [E\{B_3N_3H_{6-x}M_x\} + x\,E\{H\}] - [E\{B_3N_3H_6\} + x\,E\{M\}] \qquad (x = 1-6)$$

$$\Delta E_s = E_s[B_3N_3H_{6-x}M_x] - E_s[B_3N_3H_{6-x-1}M_{x-1}]$$

where $E\{..\}$ represents the total electronic energy of respective species including zero point correction.

## 3. Results and discussion

We start our discussion considering borazine ($B_3N_3H_{16}$) and Li atom. Borazine is planar ring system with all bond lengths equal to 1.43 Å as shown in Fig. 1. It is stabilized due to charge transfer from electropositive B to electronegative N atoms such that all B-N bonds become polar. In Table 1, we have listed calculated VIE of $B_3N_3H_6$ and Li, and compared with corresponding experimental values [30, 31]. One can note that B3LYP/aug-cc-pVDZ calculated VIE are in good agreement with the experiments. Therefore, present computational scheme is capable to provide reasonable results for the systems under study.

Next, we consider successive substitution of Li atoms on various positions of the ring. For a single Li atom, there are two possible sites, B and N of the ring. As displayed in Fig. 1, the lowest energy structure of $B_3N_3H_5Li$ corresponds to the one in which Li is substituted at N site. This can be expected due to strong interaction between electronegative N and electropositive Li, unlike $B_3N_3H_5X$ (X = OH, $NO_2$, $NH_3$, $CH_3$, $CF_3$) which possesses lower energy with X substituted at B-site [32, 33]. For substitution of two Li atoms, there are four possible isomers with relative energies 38.2−75.9 kcal/mol but in the lowest energy isomer of $B_3N_3H_4Li_2$, both Li atoms bind to N-sites. Likewise, the relative energy of six possible isomers of $B_3N_3H_3Li_3$ ranges from 34.2 kcal/mol to 116.2 kcal/mol and three Li atoms substituted at N-sites leads to minimum energy structure. When number of Li atoms ($x$) exceeds 3, the lowest energy structure of $B_3N_3H_{6-x}Li_x$ assumes the configuration so as to minimize B−Li and Li−Li repulsions. For instance, in lowest energy structure of $B_3N_3H_2Li_4$, three Li atoms are substituted at three N-sites and additional Li atom lies at B-site opposite to N-site. Similarly, $B_3N_3HLi_5$ attains minimum energy when three Li atoms occupy N-sites and remaining two occupy B-sites. For $B_3N_3HLi_5$ and $B_3N_3Li_6$ in which all H atoms are substituted with Li atoms, one can see that B−Li and N−Li are significantly rotated with respect to substitution site. The rotation of B−Li bonds can be accredited to combined effect of the repulsion between B and Li, and attraction of Li with neighboring N atom. The rotation of N−Li bonds, on the other hand, is due to repulsion between Li substituent atoms. It is due to the rotation of B−Li and N−Li bonds that a previous study [34] regarded $B_3N_3Li_6$ as a star like structure. Like $B_3N_3H_6$, the bond length equalization takes place in $B_3N_3Li_6$ such that all bond become equal to 1.45 Å, slightly larger than borazine.

The substitution energy ($E_s$) and its growth ($\Delta E_s$) for $B_3N_3H_{6-x}Li_x$ are collected in Table 2. One can note that all $B_3N_3H_{6-x}Li_x$ are kinetically stable due to $E_s > 0$. Furthermore, their kinetic stability increases as $E_s$ increases monotonically. In particular, the increment in $E_s$, i.e., $\Delta E_s$ become significantly larger (3.05 eV) when $x$ exceeds 3. The enhanced stability of $B_3N_3H_{6-x}Li_x$ ($x > 3$) might be associated with their lower VIE values. The VIE of $B_3N_3H_{6-x}Li_x$ species are also listed in Table 2 as a function of $x$. One can note that VIE decreases successively with the substitution of Li atoms. For $x > 3$, i.e., when all N-sites are substituted with Li atoms, the VIEs of $B_3N_3H_{6-x}Li_x$ become significantly smaller and lower than that of Li atom. Therefore, $B_3N_3H_{6-x}Li_x$ species behave as superalkali molecules for $x > 3$. In order to explain the trend of VIE values, we have listed NBO charges on Li ($Q_{Li}$) in Table 2. One can note that $Q_{Li}$ decreases only slightly from $0.88e$ to $0.84e$ with the increase in Li atoms up to $x = 3$. For $x = 4$, $Q_{Li}$ decreases rapidly, which leads to smaller VIE value for $x > 3$. It is also interesting to note that the VIE of $B_3N_3HLi_5$ (4.14 eV) is smaller than $B_3N_3Li_6$ (4.25 eV). This feature can also be explained on the basis of charge localization on Li atoms. Although, average $Q_{Li}$ value for $B_3N_3HLi_5$ is larger than $Q_{Li}$ of $B_3N_3Li_6$, the VIE of $B_3N_3HLi_5$ decreases due to presence of Li atom in the vicinity of B, which weakens B−H bond. This cause to destabilize the $B_3N_3HLi_5$, which is reflected in its smaller $\Delta E_s$ value, consequently its VIE value.

It is also interesting to note that all $B_3N_3H_{6-x}Li_x$ species are planar, like borazine. Note, however, that the aromaticity of borazine has still been a controversy [35]. The most reliable aromaticity measure, i.e., magnetic criterion [36] suggests that the borazine ring electrons generate a paratropic region in the ring centre (similar to an anti-aromatic response). In order to quantify it, we have calculated perpendicular tensor (ZZ) component of nucleus independent chemical shift (NICS) at the ring centre. It is known that negative $NICS_{ZZ}$ values correspond to diatropic current, whereas positive $NICS_{ZZ}$ lead to paratropic current. The calculated NICS$_{ZZ}$ value of borazine 11.62 ppm clearly suggests that the pi electrons of borazine support paratropic current, reflecting anti-aromatic feature of the ring, as suggested previously [35]. With the substitution of Li in borazine, NICS$_{ZZ}$ values of ring decrease successively. Therefore, the contribution of pi electrons in paratropic current decreases, which can be expected due to charge transfer from Li atoms. Thus, the Li substitution in the borazine tends to destroy the anti-aromatic response of the ring.

Above discussions establish that it is possible to design planar superalkali species by successive substitution of Li atoms in borazine. In order to verify that similar conclusions apply to other alkali metals, we have also optimized the structure of B$_3$N$_3$Na$_6$ as displayed in Fig. 2. One can note that the structure of B$_3$N$_3$Na$_6$ is planar with equal ring bond lengths, like B$_3$N$_3$Li$_6$. However, the ring bond lengths (1.41 Å) of B$_3$N$_3$Na$_6$ are smaller than those of B$_3$N$_3$Li$_6$ (1.45 Å). Note that NBO charges on B are 0.30$e$ for B$_3$N$_3$Li$_6$ but 0.58$e$ for B$_3$N$_3$Na$_6$ whereas N atoms in both systems possess equal charges of -1.48$e$. This suggests that B−N bonds in B$_3$N$_3$Na$_6$ are stronger than those in B$_3$N$_3$Li$_6$. This is due to larger size of Na which reduces the charge transfer to ring, consequently strengthens B−N bonds. Note that $Q_{Li}$ for B$_3$N$_3$Li$_6$ is 0.59$e$ whereas $Q_{Na}$ in B$_3$N$_3$Na$_6$ is only 0.45$e$. The VIE of B$_3$N$_3$Na$_6$ is calculated to be 3.96 eV, which is smaller than that of B$_3$N$_3$Li$_6$, and comparable to the IE of Cs, 3.89 [30] but the substitution energy (16.75 eV) is larger. Likewise, the VIE of substituted borazine can be further reduced by using K atoms.

## 4. Conclusions

We have performed B3LYP/aug-cc-pVDZ level calculations on successive Li-substitutions in borazine (B$_3$N$_3$H$_6$) at various positions. The lowest energy structures of B$_3$N$_3$H$_{6-x}$Li$_x$ are identified for $x$ = 1 to 6. It has been noticed that that the vertical ionization energy (VIE) of substituted borazine are successively reduced with the increase in Li atoms. Furthermore, the planarity of the ring is preserved but the NICS$_{zz}$ value at the ring's center is successively reduced. With $x \geq 4$, the $B_3N_3H_{6-x}Li_x$ species behave as superalkalis as their VIE becomes lower than that of Li atom. For instance, the VIE of $B_3N_3Li_6$ is reduced to 4.25 eV from 10.09 eV of $B_3N_3H_6$. We have also studied $B_3N_3Na_6$ and noticed that the VIE is further reduced to 3.96 eV. These findings suggest the alkalization of borazine as an effective route to the design of new planar superalkalis. Further studies on the properties and applications of these species are ongoing in our lab.

**Acknowledgement**

A. K. Srivastava acknowledges Council of Scientific and Industrial Research (CSIR), New Delhi, India for a research fellowship [Grant No. 09/107(0359)/2012-EMR-I].

Table 1. Comparison of B3LYP/aug-cc-pVDZ calculated values with experiments.

| Ionization energy (eV) | Li | Borazine |
| --- | --- | --- |
| Calculated VIE | 5.431 | 10.089 |
| Experimental value | 5.392[a] | 10.01±0.01[b] |

[a]Ref. [30]
[b]Ref. [31]

Table 2. Calculated parameters for the lowest energy $B_3N_3H_{6-x}Li_x$ structures

| $x$ | VIE (eV) | $E_s$ (eV) | $\Delta E_s$ (eV) | $NICS_{ZZ}$ (ppm) | $Q_{Li}$ ($e$) |
|---|---|---|---|---|---|
| 1 | 8.82 | 1.48 |  | 10.03 | 0.88 |
| 2 | 7.59 | 3.01 | 1.53 | 7.66 | 0.86 |
| 3 | 7.18 | 4.54 | 1.53 | 4.67 | 0.84 |
| 4 | 5.37 | 7.59 | 3.05 | 3.89 | 0.48[a] |
| 5 | 4.14 | 9.62 | 2.03 | 1.30 | 0.62[a] |
| 6 | 4.25 | 11.88 | 2.26 | 1.47 | 0.59 |

[a]Average charge value

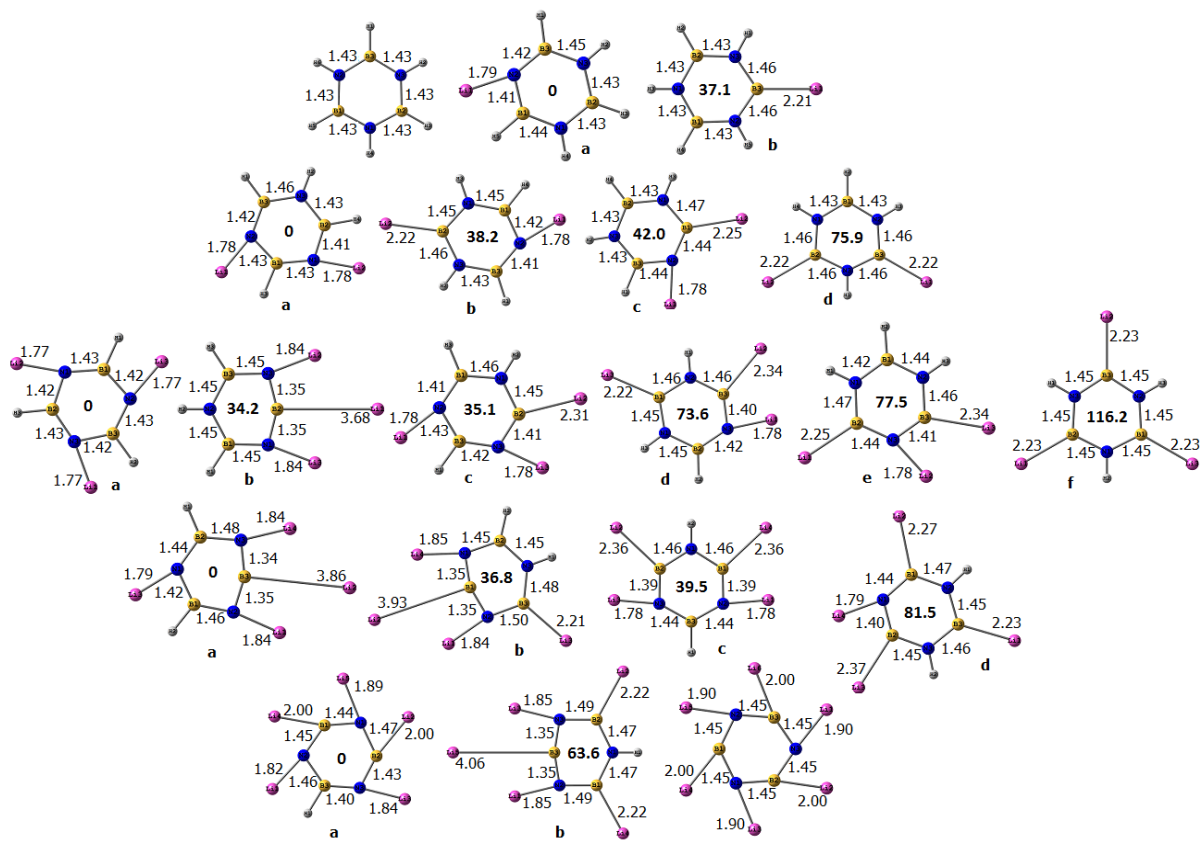

Fig. 1. Equilibrium structures of possible isomers of $B_3N_3H_{6-x}Li_x$ ($x = 0-6$) obtained at B3LYP/aug-cc-pVDZ level. Bond lengths (in Å) and relative energy (in kcal/mol) are also given.

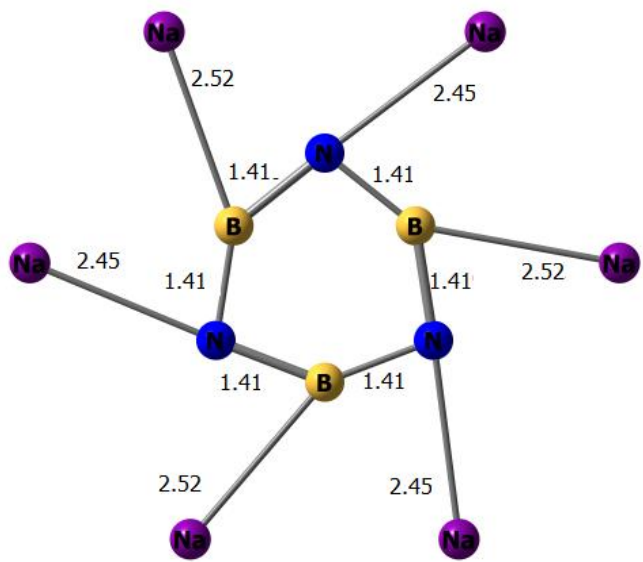

Fig. 2. Equilibrium structure of $B_3N_3Na_6$ with bond lengths in Å obtained at B3LYP/aug-cc-pVDZ level.